 \documentstyle[12pt]{article}
\begin{document}

\title{Thick domain walls in a polynomial approximation
\thanks{Paper supported
in part by the grant KBN 2 P302 049 05.}}

\author{by\\
\\
H. Arod\'z   \\
\\
Institute of Physics, Jagellonian University,
Cracow \thanks{Address: Reymonta 4, 30-059 Cracow, Poland.}
        \thanks{E-mail: ufarodz@ztc386a.if.uj.edu.pl} }

\date{  $\;\;$}
\maketitle

\thispagestyle{empty}
 $\;\;\;$ \\
{\bf Abstract} \\
Relativistic domain walls are studied in the framework of a
polynomial approximation to the field interpolating between different vacua
and forming the domain wall.  In this approach we can calculate evolution of
 a core and of a width of the domain wall.
  In the single, cubic polynomial approximation used in this paper,
   the core   obeys Nambu-Goto equation for a relativistic membrane.
   The width of the
 domain wall obeys a nonlinear equation which is solved perturbatively.
 There are two types of corrections to the constant zeroth
  order width: the ones oscillating in time, and the corrections directly
  related to curvature of the core. We find that curving a static
   domain wall is
  associated with an increase of its width.
   As an example, evolution of a toroidal domain wall is
  investigated.

$\;\;$   \nopagebreak  

\pagebreak

\setcounter{page}{1}
\section{Introduction}
Recently one observes rapidly growing interest in  time
evolution of topological defects in 3+1 dimensional space-time. This subject
is important for many branches of physics. Without attempting to
present here a complete list, let us mention vortices in
superconductors \cite{1}
 and in superfluids \cite{2}, defects in liquid crystals \cite{3},
magnetic domain walls \cite{4},  cosmic strings \cite{5}, \cite{6}, and
 a flux  tube in QCD \cite{7}.
 While evolution of topological defects in 1+1 dimensional space-time
  has been rather well understood, in 3+1 dimensions relatively
 little is known, and the problem is actually a formidable one.

 The type of equations from which one attempts to calculate evolution
 of topological defects depends in an essential manner on the physical
 context. In condensed matter physics one uses e.g. diffusion type
 equations \cite{8} or nonlinear Schr\"{o}dinger type equations \cite{2}.
 For cosmic strings in a negligible gravitational field, or for the particle
  physics flux-tubes, one should use Poincar\'{e}
 invariant wave equations.

  Our paper is devoted to dynamics of domain walls governed by a
  Poincar\'{e} invariant wave equation.
 In this case, several analytical approaches have been made, see
 e.g.\cite{9},\cite{10},\cite{11},\cite{12} as well as numerical calculations,
 see, e.g. \cite{13},\cite{14}. They have given  insights into the
 dynamics of domain walls, and have revealed the richness and intricacy  of
it. Our motivation for studying the dynamics of relativistic domain walls
is of rather mathematical character -- relative simplicity of the pertinent
field equations
makes them a convenient testing ground for new methods of calculating the
evolution of topological defects. Nevertheless, direct physical applications
are also possible. Relativistic domain walls appear in a field-theoretic
approach to cosmology, and in particle physics (e.g. the surface of a quark
bag can be regarded as a domain wall). Moreover, the polynomial approximation
we develop in this paper can also be applied to non-relativistic domain walls
observed in condensed matter physics - no significant changes are required.

Till now, the main line of the analytical approaches to the description
of dynamics of a
domain wall has been to reduce the initial, 3+1 dimensional field-theoretical
system to an effective theory of a classical, relativistic membrane.
This approach, called the effective action method, is very appealing
conceptually,
 and it is correct in principle. However, because of the complexity of the
 pertinent field-theoretic equations, it is rather difficult to carry out the
 necessary calculations without shortcuts, which in turn introduce some
 uncertainty about the final result. For a critical
 discussion of the effective action method  we refer
 the reader to the paper \cite{15}.
 For the most recent application of the effective action method to domain
 walls, see \cite{16}.

 Our opinion is that at the present stage of the subject
  one should develop and refine various
 methods of investigations of the dynamics of the domain walls. It seems that
 because of enormous complexity of the full, 3+1 dimensional, nonlinear
  field-theoretic dynamics
 involved, it is long time until we can analytically or numerically
 calculate evolution of a generic domain wall without any difficulty.

In the present paper we generalize the method of analysis of evolution
 of  relativistic domain walls proposed in \cite{17}.
 The main characteristic
 feature of this approach  is a simple,
 approximate  polynomial Ansatz for the field inside the domain wall, while
 outside of it the field has the exact vacuum values.
The coefficients of this polynomial are calculated from the field equation and
from boundary conditions. Actually, this
 approximation can be regarded as an application of splines, \cite{18}.
 In the paper \cite{17} this approach has been applied only to
 cylindrical and spherical domain walls.  In the present paper we apply
  the polynomial approximation to generic smooth domain walls - this is
 presented in Section 2. We approximate the field inside the domain wall
 by a cubic polynomial in a transverse, co-moving coordinate.
  We obtain Nambu-Goto type equation for a core
 of the domain wall,
and a nonlinear equation for the width of the domain wall. Also, a
perturbative scheme for solving the latter equation is presented. It is
based on dividing the width into two components: the one oscillating
with characteristic frequency given by the mass of the scalar field, and the
other one directly related to curvature of the core.

  In the next Section we rewrite the Nambu-Goto
 equation, and a formula for the width of the domain wall,
  in terms of local curvature radia and velocity perpendicular
 to the core. This gives a rather nice and useful insight into the
 dynamics of the domain wall. We show that a force acting on a small piece
 of the core can be regarded as being due to a surface tension. We also
 show that curving a static domain wall is associated with an increase of
 its width.

In Section 4 we calculate the energy of the domain wall. We find that the
energy density depends on the curvature of the core, and that
 nonuniformities of the width increase the energy density. We also find,
 rather surprisingly, that curving the domain wall seems to decrease
  its energy.  These results, as well as all others in this paper, are
   obtained for slightly
 curved domain walls; our approximations break down when the curvature radia
become comparable with the width of the domain wall.

  Next, in Section 5, we consider as an example evolution of a
   toroidal domain wall. We
 find two types of evolution of this
 domain wall.

  Section 6 contains a suggestion how to improve
 our approximate Ansatz for the scalar field, and also some ending remarks.

 In  the Appendix  we discuss accuracy of the polynomial approximation.

\section{General formalism for evolution of the domain
 wall in the polynomial approximation}

In this subsection we would like to generalize the formalism presented in
\cite{17}, where only cylindrical and spherical domain walls were
considered, and to present the corresponding, approximate solution of
 the field equation.

We will investigate domain walls in the well-known model, see e.g. \cite{6},
involving only a  single, real
scalar field $\Phi$ with the following
Lagrangian
\begin{equation}
 L=-\frac{1}{2}\eta_{\mu\nu}\partial^{\mu}\Phi \partial^{\nu}\Phi
-\frac{\lambda}{2}(\Phi^{2}-\frac{M^2}{4\lambda})^2,   \end{equation}
where $(\eta_{\mu\nu})=diag(-1,1,1,1)$, and $\lambda,M$ are positive
constants. The corresponding field equation is
\begin{equation} \partial_{\mu}\partial^{\mu}\Phi -
2 \lambda (\Phi^2-\frac{M^2}{4\lambda})\Phi=0. \end{equation}

The vacuum values of the field $\Phi$  are equal to $\pm \Phi_0$,
where $\Phi_0 \equiv M/2\sqrt{\lambda}$. The domain wall arises
if at a given time the field is equal to  one of the two vacuum values
 in some region of
the space, and is equal to the other vacuum value
in the complementary part of the space,
except for the border layer between the two regions
(the domain wall), where the field
smoothly interpolates between the vacuum values. It is clear that at each
instant of time the field $\Phi$
vanishes somewhere  inside the border layer. We assume that the locus of
these zeros is a smooth surface $S$.
 We shall call it the core of the domain wall. The
well-known example of the domain wall, with the static core given by the
$(x^1,x^2)$ plane, is given by the following exact, static solution of Eq.(2)
\begin{equation}
\Phi = \Phi_0\; tanh(\frac{x^3}{2l_0}),
\end{equation}
where $l_{0} \equiv M^{-1}.$  The width of this domain wall is of the order
$l_{0}$, and energy density is exponentially localised around the $(x^1,x^2)$
plane.

For a  generic domain wall, space-time parametrisation of the
 world-volume $\Sigma$
of the core (a 3-dimensional manifold embedded in Minkowski space-time,
whose time slices coincide with $S$) can be chosen as follows
\begin{equation}
\left( X^{\mu}\right) (\tau,\sigma^1,\sigma^2) = \left(  \begin{array}{c}
\tau \\  X^1(\tau,\sigma^1,\sigma^2) \\ X^2(\tau,\sigma^1,\sigma^2) \\
X^3(\tau,\sigma^1,\sigma^2)
\end{array} \right),
\end{equation}
where $\tau$ coincides with the laboratory frame time $x^{0}$, and $\sigma^1,
\sigma^2$ parametrise the core $S$ at each  instant of time.

As usual, we introduce a special coordinate system
$(\tau,\sigma^1,\sigma^2,\xi)$
in a vicinity of the world-volume $\Sigma$, co-moving with the core,
\cite{9}.  The new coordinates
$(\tau, \sigma^1, \sigma^2, \xi)$ are defined by the
following formula
\begin{equation}
x^{\mu} = X^{\mu}(\tau,\sigma^1,\sigma^2) + \xi n^{\mu}(\tau,\sigma^1,
\sigma^2),
\end{equation}
where $x^{\mu}$ are Cartesian lab-frame coordinates in Minkowski
space-time, and $(n^{\mu})$ is a normalised space-like four-vector,
 orthogonal
to the $\Sigma$ (in the covariant sense) i.e.
\begin{equation}
n_{\mu}X^{\mu}_{,a} = 0,  \;\;\;\;\; n_{\mu}n^{\mu}=1,
\end{equation}
where $a=0$ corresponds to $\tau$; $a=1, a=2$ correspond
to $\sigma^1,\sigma^2$,
and $X^{\mu}_{,\tau}\equiv \partial X^{\mu}/
\partial\tau$, etc.  The four-vectors $X_{,\tau}, X_{,\sigma^1}, X_{,\sigma^2}$
are tangent to $\Sigma$. For points lying on the core $\xi = 0$,
and the parameter $\tau$ coincides with the lab-frame time $x^{0}$.
For $\xi \neq 0$  $\tau$ is not equal to the lab-frame time $x^0$.
  The advantage of using the co-moving coordinates is that the world-volume
  $\Sigma$   is
described by the simple condition $\xi=0$. Notice that the definition (5)
implies that $\xi$ is a Lorentz scalar.

The next step is to write Eq.(2) in the new coordinates. It is convenient
to introduce extrinsic curvature coefficients $K_{ab}$ and induced
metrics $g_{ab}$ on $\Sigma$:
\begin{equation}
K_{ab} = n_{\mu} X^{\mu}_{,ab}, \;\;\;\;  g_{ab}=X^{\mu}_{,a} X_{\mu,b},
\end{equation}
where $a,b=0,1,2$. The covariant metric tensor in the new
coordinates can be readily calculated, and it can be
written in the following form
\begin{equation}
[G_{\alpha\beta}] = \left[ \begin{array}{lr} G_{ab} & 0 \\
0 & 1 \end{array} \right],
\end{equation}
where $\alpha,\beta=0,1,2,3$;\  $\;\;\;\alpha=3$ corresponds to the $\xi$
coordinate; and
\begin{equation}
G_{ab}=M_{ac} g^{cd} M_{db},\;\;\; M_{ac}\equiv g_{ac} - \xi K_{ac}.
\end{equation}
Thus, $G_{\xi\xi}=1, \; G_{\xi a}=0$ (a=0,1,2, as in (6)).
It follows from formula (9) that $\sqrt{-G}$, where
$G\equiv det[G_{\alpha\beta} ]$, is given by the formula
\begin{equation}
\sqrt{-G}=-(\sqrt{-g})^{-1} \; det M,
\end{equation}
where as usual  $g\equiv det[g_{ab}]$.
$det M$ can be explicitely evaluated in terms of $K_{ab}$ and $g_{ab}$:
\begin{equation}
det M = h(\tau,\sigma^1,\sigma^2,\xi) \;\;g(\tau,\sigma^1,\sigma^2),
\nonumber
\end{equation}
where
\begin{equation}
h(\tau,\sigma^1,\sigma^2,\xi) =
1 - \xi K^{a}_{a} +\frac{1}{2} \xi^2 (K^{a}_{a} K^{b}_{b} - K^{b}_{a}
K^{a}_{b}) -\frac{1}{3!} \xi^3 \epsilon_{abc}\epsilon^{def} K^{a}_{d} K^{b}_{e}
K^{c}_{f}.             \nonumber
\end{equation}
In formula (11) we have noted that $g$ can depend on $\tau,\sigma^1,\sigma^2$.
Thus,
\begin{equation}
\sqrt{-G}= \sqrt{-g}\; h.
\end{equation}
For raising and lowering the latin indices of the extrinsic
curvature coefficients we use the induced metric tensors
$g_{ab},\;\; g^{ab}$.
The inverse metric tensor $G^{\alpha \beta}$ is given by
\begin{equation}
[G^{\alpha \beta}] = \left[ \begin{array}{lr} G^{ab} &  0 \\
0 &   1 \end{array} \right],
\end{equation}
where
\begin{equation}
G^{ab}= (M^{-1})^{ac} g_{cd} (M^{-1})^{db}.
\end{equation}
 Simple algebraic calculation gives   explicit
formula for $(M^{-1})^{ac}$:
\begin{eqnarray}
(M^{-1})^{ac} = \frac{1}{h} \left\{ g^{ac} [1- \xi K^{b}_{b} +\frac{1}{2}
\xi^2 ( K^{b}_{b} K^{d}_{d} - K^{d}_{b} K^{b}_{d} )] \right. \nonumber \\
  +   \left. \xi (1-\xi K^{b}_{b}) K^{ac} + \xi^2 K^{a}_{d} K^{dc} \right\}
 \end{eqnarray}
(this is just the matrix inverse to $[M_{ab}]$; by definition it has
upper indices).
 In general, the coordinates $(\tau, \sigma^1, \sigma^2, \xi)$ are defined
locally,
 in a vicinity of the world-volume $\Sigma$.
The allowed range of the $\xi$ coordinate can be determined from the
condition $h > 0$. Detailed discussion of the region of validity of the
co-moving coordinates has been given in \cite{17}.
 In the co-moving coordinates the field equation (2) has the following form
 \begin{equation}
  \frac{1}{\sqrt{-G}}\frac{\partial}{\partial u^{a}}
(\sqrt{-G}\;G^{ab}
\frac{\partial \Phi}{\partial u^{b}})
  + \frac{1}{h} \partial_{\xi}( h \;\partial_{\xi}\Phi)
  -2 \lambda  (\Phi^2 -\frac{M^2}{4\lambda} ) \Phi =0.
 \end{equation}

 The basic feature of our approach is
 the approximate cubic polynomial Ansatz for the $\Phi$ field inside
  the domain wall. We assume that
 \begin{equation}
 \Phi(\tau,\sigma^1,\sigma^2,\xi) = \left\{ \begin{array}{lcl} +\Phi_0
  &\mbox{for} &
 \xi \geq \xi_{0}, \\
 A \xi + \frac{1}{2} B \xi^2 + \frac{1}{3!} C \xi^3  &\mbox{for}  & -\xi_{1}
\leq \xi
  \leq \xi_{0},  \\
  -\Phi_0 &\mbox{for}  & \xi \leq -\xi_{1}.     \end{array}  \right.
  \end{equation}
  Here $\xi_{0}, \xi_{1}, A, B$   and $ C$ are as yet unknown
  functions of
  $\tau, \sigma^1$ and $\sigma^2$. We assume that $\xi_{0}, -\xi_{1}$ lie
  in the allowed range of the $\xi$ coordinate - roughly speaking this is
  true when the width of the domain wall is small in comparison with  radia
  of curvature of the core in the local rest-frame of the considered piece
  of the core.

  Notice that $\pm \Phi_0$ are exact solutions of Eq.(2). They are
  defined also outside  of the region of validity of the co-moving
  coordinates. Therefore, formula (18) actually defines the field $\Phi$ in
the  whole space-time.

  The Ansatz (18) implies that in fact we introduce boundaries of the
  domain wall. The outer boundary $\vec{X}_{(+)}$ in the co-moving reference
  frame is defined by the formula
 \[ \vec{X}_{(+)}(\tau, \sigma^1, \sigma^2) =  \vec{X}(\tau, \sigma^1,
\sigma^2)
 + \xi_{0}(\tau, \sigma^1, \sigma^2) \;     \vec{n}(\tau, \sigma^1, \sigma^2),
 \]
while for the inner one ($\vec{X}_{(-)}$)
\[
\vec{X}_{(-)}(\tau, \sigma^1, \sigma^2) = \vec{X}(\tau, \sigma^1, \sigma^2) -
 \xi_{1}(\tau, \sigma^1, \sigma^2) \;  \vec{n}(\tau, \sigma^1, \sigma^2).
 \]
 Here $\vec{n}$ is the spatial part of the four-vector $(n^{\mu})$.

  Inserting the cubic polynomial  into Eq.(17) and equating to zero
  terms proportional to the zeroth and first powers of $\xi$
we obtain the following recurrence relations:
\begin{equation}
B = A K^{a}_{a},
\end{equation}
\begin{equation}
C= - \Box^{(3)} A + ( K^{a}_{b}K^{b}_{a} - 2 \lambda \Phi_{0}^2 ) A + K^{a}_{a}
B,
\end{equation}
where
\begin{equation}
\Box^{(3)} \equiv \frac{1}{\sqrt{-g}} \frac{\partial}{\partial u^{a}}(\sqrt{-g}
g^{ab}\frac{\partial}{\partial u^{b}})   \nonumber
\end{equation}
is the three-dimensional d'Alembertian on the
 world-volume $\Sigma$ of the core.
Of course, the cubic polynomial (18) does not
 obey Eq.(17) exactly. The leftover
terms in Eq.(17) are of the order $\xi^2$ and higher.
 We assume that these terms are not important. We will discuss the problem
 of accuracy of our approximation
  in the Appendix.

 We also require that the field $\Phi$ is continuous everywhere, in particular
at
the boundaries, i.e. for
  $\xi = \xi_{0}, \xi = -\xi_{1}$. Then, by a standard reasoning, we deduce
from Eq.(17) that also $\partial_{\xi}\Phi$ is continuous at the boundaries.
 The second derivative $\partial^{2}_{\xi} \Phi$ is not
continuous at the boundary, in general.
The conditions of continuity of $\Phi$  and  $\partial_{\xi}\Phi$
at $\xi=\xi_{0}$ and $\xi=-\xi_{1}$
give
\begin{equation}
\xi_{0} = \xi_{1},
\end{equation}
\begin{equation}
A = \frac{3}{2}\frac{\Phi_0}{\xi_{0}},\;\;\; B=0, \;\;\; C=-\frac{3
\Phi_0}{\xi_{0}^{3}}.
\end{equation}

The condition $B = 0 $ together with relation (19) gives
the equation of motion for the core $S$:
\begin{equation}
K^{a}_{a} = 0.
\end{equation}
It can  be shown that this equation is equivalent to Nambu-Goto equation
\begin{equation}
 \Box^{(3)} X^{\mu} = 0
 \end{equation}
 for a relativistic membrane. Thus, in the approximations we have made,
  the core can be regarded as the Nambu-Goto type relativistic membrane.

 Relations (20), (23) give also the following 2+1 dimensional,
  nonlinear wave equation
 \begin{equation}
 \Box^{(3)} \tilde{A} + (\frac{1}{2l_{0}^{2}} - K^{b}_{a} K^{a}_{b}) \tilde{A}
 -\frac{1}{ 2l_{0}^{2}} \tilde{A}^3 = 0
 \end{equation}
 for $ \tilde{A}(\tau,\sigma^1, \sigma^2) \equiv \frac{2l_{0}}{\xi_{0}}$.
$\tilde{A}$ is proportional to the inverse width of the domain wall
($2\xi_{0}$)
 measured in the natural length
 unit $l_{0}$, and  it can be regarded as a
 scalar field defined on the core of the domain wall.
 Let us note here that $2\xi_{0}$ is the width
 in the co-moving coordinates. Its transformation to the lab-frame
 is not trivial: it includes Lorentz contraction and other changes.
 This transformation has been discussed in detail in the paper
 \cite{17} in the case of cylindrical domain wall.

 Eq.(26) can be written in terms of dimensionless variables: introducing
 dimensionless $\tilde{\tau} \equiv \tau / l_{0}$ and $ \tilde{\sigma}^{1,2}
  \equiv \sigma^{1,2} / l_{0}$ , $ \tilde{\partial_{0}} \equiv \partial /
   \partial \tilde{\tau}$, etc., we have
   \begin{equation}
   \Box^{(3)} \tilde{A} + (\frac{1}{2} - l_{0}^{2} K^{b}_{a} K^{a}_{b})
\tilde{A}
   - \frac{1}{2} \tilde{A}^3 = 0.
   \end{equation}
   Because $g^{ab}$ as well as $l_{0}^{2} K^{a}_{b} K^{b}_{a}$ are
   dimensionless, all coefficients  and variables in Eq.(27) are
   dimensionless.

   Certain solutions of equation (27) for $\tilde{A}$ can be found in an
   approximation scheme which is a generalization of a perturbative
   expansion proposed
   in \cite{17} in the case of cylindrical and spherical domain walls.
In that paper one can find a detailed motivation for the subsequent steps; in
the present paper we will only briefly describe the scheme.
   Our scheme probably gives only a certain class of solutions - there
    might be
    other solutions which can not be obtained in this way. Roughly speaking,
    the idea is  to restrict our considerations to the cases such that
      the dimensionless extrinsic curvature $l_{0}^{2}
  K_{b}^{a} K_{a}^{b} $ is close to zero, and to expand $\tilde{A}$ in
  non-negative powers of it. In the zeroth order approximation Eq.(27) has
  the constant solution
  \begin{equation}
  \tilde{A}^{(0)} = 1.
  \end{equation}
There are also other solutions  in this order, e.g.,
 the ones having form of
 waves propagating along the core. Solutions of this latter type
are time-dependent and have a characteristic frequency of oscillations
 $\geq 1$, where 1 is
the (dimensionless) mass of the $\tilde{A}$ field. Notice that this mass
coincides with
the mass of the $\Phi$ field in the unit $l_{0}^{-1}$.  We assume also that
$l_{0}^{2} K^{b}_{a} K^{a}_{b}$ and its derivatives are smooth functions
 of $\tilde{\tau}, \tilde{\sigma}^1,
\tilde{\sigma}^2$,  and such that  the derivatives
$\tilde{\partial}_{b}(l_{0}^{2} K_{a}^{b} K^{a}_{b})$ are much smaller
than the function
$l_{0}^{2} K_{a}^{b} K^{a}_{b}$ itself. In this case the term $l_{0}^{2}
K_{a}^{b} K_{b}^{a} \tilde{A}$ present in Eq.(27) generates a smooth,
non-oscillating component $N$ in $\tilde{A}$.
 We introduce the following perturbative Ansatz for
$\tilde{A}$:
\begin{equation}
\tilde{A} = 1+ \Omega + N,
\end{equation}
where $\Omega$ denotes the oscillating component -- the amplitude of these
oscillations is assumed to be of the order $l_{0}^{2}
K_{a}^{b} K_{b}^{a}$ . Equation (27) can be split
into two equations:
\begin{equation}
\Box^{(3)}\Omega =  (1 + l_{0}^{2} K^{b}_{a} K_{b}^{a} + 3 N +\frac{3}{2} N^2)
 \Omega
+ \frac{3}{2} N \Omega^2  + \frac{3}{2} \Omega^2 + \frac{1}{2} \Omega^3 ,
\end{equation}
\begin{equation}
 N = \Box^{(3)} N - l_{0}^{2} K_{a}^{b} K^{a}_{b} - l_{0}^{2} K_{a}^{b}
K^{a}_{b}
 N - \frac{3}{2} N^2 -\frac{1}{2} N^3.
 \end{equation}

 The term $\Box^{(3)} N$ is by assumption regarded as small in comparison
 with the zeroth order contribution to $N$. Therefore,
  eq.(31) implies the following
result for the lowest order contribution $N^{(1)}$ to $N$:
 \begin{equation}
  N^{(1)} = - l_{0}^{2} K_{a}^{b} K_{b}^{a}.
 \end{equation}

 From Eq.(30) we obtain the equation for the zeroth order  oscillating
  contribution
 $\Omega^{(1)}$ to $\Omega$:
 \begin{equation}
 \Box^{(3)} \Omega^{(1)} =  \Omega^{(1)}.
 \end{equation}
 Detailed form of this equation depends on the metric $g_{ab}$.
 Because the frequency of oscillations implied by Eq.(33) coincides
 with the mass of the scalar field (measured in the units $l_{0}^{-1}$),
 the presence of these oscillations is quite natural. The
 domain wall with the oscillating width can be regarded as a perturbation of
 a proper domain wall, which by definition does not have the oscillating
 component.

\section{Dynamics of the core in terms of local  curvatures}

Equation (24), which governs  evolution of the core of the domain wall,
has rather abstract form. In order to bring its contents to surface, we will
write it in appropriately chosen local coordinates $\tau, \sigma^1, \sigma^2$
on the world-volume $\Sigma$ of the core $S$. We shall call them  the
physical coordinates.

As the $\tau$ coordinate we take
the one present in formula (4), without any further specification, while
the coordinates $\tau, \sigma^1, \sigma^2$ are now defined in a vicinity
 of an arbitrarily chosen point
$\vec{X}_{0}(\tau)$ of the core $S$ in a special manner explained below.
For $\vec{X}\equiv (X^1,X^2,X^3)$  in formula (4) we write
\begin{eqnarray}
\vec{X}(\tau,\sigma^1,\sigma^2) = \vec{X}_{0}(\tau) + \vec{e}_{1}(\tau)
\sigma^1
+ \vec{e}_{2}(\tau) \sigma^2  \nonumber    \\
+\frac{1}{2} \vec{e}_{1}(\tau)\times
 \vec{e}_{2}(\tau) \left[\frac{(\sigma^1)^2}{R_{1}(\tau)} +
 \frac{(\sigma^2)^2}{R_{2}(\tau)} \right] + {\cal O}^{(3)}(\sigma),
 \end{eqnarray}
where $\times$ denotes the vector product, and $\vec{e}_1, \;\vec{e}_{2}$ are
unit vectors specified below. For our purposes, the terms
${\cal O}^{(3)}(\sigma)$ (third order terms in $\sigma^1, \sigma^2$) do
not have to be written explicitely.
It is clear that $\sigma^1 = 0 = \sigma^2 $ corresponds to the point
$\vec{X}_{0}
(\tau)$, and that the vectors $\vec{X}_{,\sigma^1}, \vec{X}_{,\sigma^2}$
tangent to $S$ at $\vec{X}_{0}(\tau)$ are equal to $\vec{e}_{1}(\tau),
\vec{e}_{2}(\tau),$ correspondingly. From conditions (6) we  obtain
that at $\vec{X}_{0}(\tau)$
\begin{equation}
n^{0} = \frac{\vec{m} \dot{\vec{X}}_{0}}{\sqrt{1 - (\dot{\vec{X}}_{0}
\vec{m})^2}},
\;\;\; \vec{n} = \frac{\vec{m}}{\sqrt{1-(\vec{m}\dot{\vec{X}}_{0})^2}},
\end{equation}
where
\begin{equation}
              \vec{m} \equiv \vec{e}_{1}\times \vec{e}_{2}
\end{equation}
is the unit vector normal to $S$ at $\vec{X}_{0}(\tau)$
and $\dot{\vec{X}}_{0} \equiv d\vec{X}_{0}/d\tau$. The extrinsic
curvature coefficients at $\vec{X}_{0}(\tau)$ are equal to
\[
K_{00} = \frac{\ddot{\vec{X}}_0 \vec{m}}{\sqrt{1-
(\vec{m}\dot{\vec{X}}_0)^2}}, \;\;
K_{0i} = K_{i0} =  \frac{\dot{\vec{e}}_i \vec{m}}
{\sqrt{1-(\vec{m}\dot{\vec{X}}_0)^2}},
\]
\begin{equation}
K_{12}=0, \;\;\; K_{11} = \frac{1}{\sqrt{1-(\vec{m}\dot{\vec{X}}_0)^2}}
\frac{1}{R_{1}(\tau)},
 \;\;\; K_{22} = \frac{1}{\sqrt{1-(\vec{m}\dot{\vec{X}}_0)^2}}
\frac{1}{R_{2}(\tau)},
\end{equation}
as it follows from the definition (7).
The fact that $K_{12}$ vanishes means that the tangent
vectors $\vec{e}_{1}, \vec{e}_{2}$ have been
chosen in directions tangent to the circles of main curvatures of the
surface S at the point $\vec{X}_{0}(\tau)$. The parametrisation (34) is the
most natural local parametrisation of the core from the viewpoint of
geometry of surfaces in the 3-dimensional space.

 As for the choice of $\vec{X}_{0}(\tau)$ at different times $\tau$, there is
 a large freedom due to the reparametrisation invariance of the equation
  (24). In  physical terms, this invariance means
  that the core $S$ is a "structureless"
   surface in the sense that translations along $S$ are not observable.
 Natural choice is that $\vec{X}_{0}(\tau)$ is a smooth
function of $\tau$. Apart from this,
 $\vec{X}_{0}(\tau)$ for different $\tau$'s can be chosen almost
arbitrarily. Even the condition $\dot{\vec{X}}^{2} < 1$ does not have
to be imposed,
because the core of the domain wall is merely a mathematical construct; in
fact there is a numerical evidence that in a related case of vortices the core
can move with superluminal velocity, see e.g.\cite{19}. We will use this large
freedom to choose $\vec{X}_{0}(\tau)$ for different $\tau$ in such a manner
that
\begin{equation}
\dot{\vec{X}}_{0}(\tau) \vec{e}_{1}(\tau) = \dot{\vec{X}}_{0}(\tau)
\vec{e}_{2}(\tau)
 = 0,
 \end{equation}
 i.e. the motion of the point $\vec{X}_{0}(\tau)$ is always in the
 direction perpendicular to $S$.
 Taking (38) into account, we can write
 \begin{equation}
 \dot{\vec{X}}_{0}(\tau) = v(\tau) \vec{m}.
 \end{equation}
 Then, simple
 calculations show that
 Eq.(24) reduces to the following relation
 \begin{equation}
 \frac{\dot{v}(\tau)}{1 - v^{2}(\tau)} = \frac{1}{R_{1}(\tau)}
 + \frac{1}{R_{2}(\tau)}.
\end{equation}
Thus, locally, the acceleration $\dot{v}$ of the piece of the core is
determined by the main radia of
curvature, up to the  Lorentz factor.

In the particular cases of a sphere ($R_{1}=R_{2}\equiv R,\; \dot{v} = -
\dot{R}$),
and of a cylinder ($R_{1}=\infty,\; R_{2} = R,\; \dot{v} = - \dot{R}$), the
usual
spherical or cylindrical coordinates, respectively, have the properties
required by formula (34) and conditions (38), and (40) coincides with
equations considered in \cite{17}.  In a more general case, the explicit
constructing of the special coordinates $(\sigma^1, \sigma^2)$,
 such that formula (34) and
condition (38) hold, might be difficult. Nevertheless, formula (40) is
very helpful in qualitative analysis of motion of the core.

If we drop the gauge condition (38) while still keeping  (34), then instead of
formula (40) we obtain
\begin{eqnarray}
\lefteqn{\dot{v}  - (\vec{m} \dot{\vec{e_{1}}}) (\dot{\vec{X_{0}}}
\vec{e_{1}})
   -(\vec{m} \dot{\vec{e_{2}}}) (\dot{\vec{X_{0}}} \vec{e_{2}}) = } \nonumber
\\
& & [1 - (\dot{\vec{X_{0}}})^{2} + (\dot{\vec{X_{0}}} \vec{e_{2}})^{2}]
\frac{1}{R_{1}}  + [1 - (\dot{\vec{X_{0}}})^{2} + (\dot{\vec{X_{0}}}
\vec{e_{1}})^{2}] \frac{1}{R_{2}},
\end{eqnarray}
where $v \equiv \dot{\vec{X_{0}}} \vec{m}$ is the perpendicular
 component of the
velocity $\dot{\vec{X_{0}}}$.

Formula (40) can be reinterpreted in terms of a surface tension.
 To show this we pass to the rest-frame of a small piece $dS$ of the core.
 (Observe that formula (40) contains the laboratory frame quantities: the
 time $x^0 = \tau$, the transverse velocity
 $v \equiv \dot{\vec{X}}_0\vec{m}$, and the radia
 $R_1, \;R_2.$) The core element $dS$ is
  parametrised by the parameters $\sigma^1, \sigma^2$ introduced by
  formula (34): $\sigma^1=0=\sigma^2$ corresponds to the "center"
  $\vec{X}_{0}(\tau)$ of  the element $dS$ of the core.
 The radia $R_1, R_2$ are now the rest-frame
  curvature radia, denoted by $R_1^{rest}, R_2^{rest}$.
 For simplicity, we restrict   our considerations to the case
   when the piece $dS$ of the core neither rotates
 nor is deformed at the chosen instant of time.
 Then $\dot{\vec{e_1}} = \dot{\vec{e_2}}= 0$ and $\dot{\vec{m}}=0$.
   Let us assume   that on a piece $\vec{dl}$ of the boundary
 $\partial(dS)$ of $dS$ acts a force $d\vec{F}$ of the magnitude
 $\omega \; dl$ ($dl\equiv |\vec{dl}|$), tangent to the core and
 perpendicular to $\vec{dl}$, directed to the outside of $dS$. Here
  $\omega$ is a constant.
  It is easy to
  see that
  \[d\vec{F}\;=\; \omega\; \vec{N}(\sigma^1,\sigma^2) \times \vec{dl}, \]
  where
  \[ \vec{N}(\sigma^1, \sigma^2) \equiv \frac{\vec{X}^{rest}_{,\sigma^{1}}
   \times
  \vec{X}^{rest}_{,\sigma^{2}}}{|\vec{X}^{rest}_{,\sigma^{1}}
   \times \vec{X}^{rest}_{,\sigma^{2}}|}
  \]
  is a unit vector perpendicular to the core at $\vec{dl}$. In the
   last formula, the vectors
   $\vec{X}^{rest}_{,\sigma^{1}},\;\; \vec{X}^{rest}_{,\sigma^{2}}$ are
    the vectors tangent
   to $S$ at the point $\vec{X}(\tau, \sigma^1, \sigma^2)$,
   calculated from  the rest frame counterpart of formula (34), i.e.
    \begin{equation}
 \vec{X}^{rest}_{,\sigma^{1}} = \vec{e}_{1} +\frac{\sigma^1}{R_{1}^{rest}}
 \vec{m},  \;\;\;
 \vec{X}^{rest}_{,\sigma^{2}} = \vec{e}_{2} +\frac{\sigma^2}{R_{2}^{rest}}
 \vec{m}.   \nonumber
 \end{equation}
 Notice that $\vec{e}^{rest}_{i} = \vec{e}_{i}$ -- the vectors tangent
 to $dS$ at $\vec{X}_{0}(\tau)$  --  do
 not change under the boost in the perpendicular direction given by
$\vec{m}$. The same is true for the parameters
$\sigma^1, \sigma^2$.  Next step is to
   calculate the integral
   \[ \int_{\partial(dS)} \vec{dF},  \]
   which gives
 the total force acting on $dS $. In the limit $dS$ shrinking
   to the point $\vec{X}_{0}(\tau)$, the result for the integral is
   \[ \vec{f}_{rest} =  \omega (\frac{1}{R_{1}^{rest}} +
    \frac{1}{R_{2}^{rest}}) |dS| \vec{m}, \]
    where $|dS|$ denotes the area of $dS$. (In fact, it
     is the dominating contribution only;
we have neglected terms which vanish faster than $|dS|$.)
  In order to find the corresponding
    force in the laboratory frame, we apply the boost with velocity
    $v \equiv |d\vec{X}_{0}/dt|$ in the direction
      $\vec{m}$: $\vec{f}_{lab} = (\sqrt{1-v^2})^{-1/2} \vec{f}_{rest}$, i.e.
    \[ \vec{f}_{lab} =  \omega |dS| \frac{1}{\sqrt{1-v^2}}
     (\frac{1}{R_1^{rest}} +
    \frac{1}{R_2^{rest}})\; \vec{m}. \]
   The  laboratory frame curvature radia
   are related to the rest frame ones by the formula
    \[ R_{i} =\;\frac{ R_{i}^{rest}}{\sqrt{1- v^2}} , \;\;\;\;\; i=1,2. \]
    Notice that
    $R_{i}$ are bigger than $R_{i}^{rest}$ -- this is due to the fact that
in the laboratory frame
    the surface $S$ is flattened in the direction of motion by  Lorentz
    contraction. Another way to obtain this transformation law is
    to use the formulae (37) and the fact that $K_{11}, K_{22}$
    at $\sigma^1 = \sigma^2 = 0$ are
    invariant under Lorentz boosts in the $\vec{m}$ direction.
       Relativistic Newton equation of motion in the laboratory frame has the
    following form
     \begin{equation}
     \omega |dS| \frac{d^{2}\vec{X}_{0}}{ds^{2}} =
     \frac{\omega |dS|}{1-v^2} (\frac{1}{R_1} + \frac{1}{R_2})
     \vec{m},
     \end{equation}
  where $s$ denotes the proper time. We have assumed that the rest-frame
 "mass" of the piece of the core comes entirely from the surface tension
 and is equal to $\omega |dS|$. Equation (43) reduces to relation (40): one
 should relate the $s$ variable to the laboratory time $\tau$ $(ds =
 \sqrt{1-v^2} d\tau)$, to substitute $d\vec{X}_{0}/ds =
 v\; \vec{m}/\sqrt{1-v^2}$,
 and to use the assumption $\dot{\vec{m}}=0$.
If the piece $dS$ rotates or is subjected to a
 deformation, the above presented
reasoning should be generalized by introducing an appropriate
stress tensor for the core. We will not dwell on this.

 At the end of the previous Section we have obtained the first correction
 to the  inverse dimensionless width of the domain wall in the co-moving frame.
  In the absence
 of the oscillatory component it is equal to
  $N^{(1)}$ given by formula (32). The
 quantity $K^{a}_{b}K^{b}_{a}$ present in that formula can easily
 be calculated in the physical coordinates $(\sigma^1, \sigma^2)$
  defined by formula (34). We
 do not assume here the gauge (38). We obtain the following result:
  at the point $\vec{X}_{0}(\tau)$
  \begin{equation}
 \frac{\xi_{0}}{2 l_0} \approx 1 +
 l_0^2 \;K^{a}_{b}K^{b}_{a}, \nonumber
 \end{equation}
 where
 \begin{eqnarray}
    \lefteqn{K^{a}_{b}K^{b}_{a} =
    \frac{2}{1-v^2} \left(\frac{1}{R_1^2} +
 \frac{1}{R_2^2} +\frac{1}{R_{1} R_{2}} \right) } \\
     &  &                   - \frac{2}{(1-v^2)^2}
  \left[(\dot{m}_1+\frac{v_1}{R_1})^2 +(\dot{m}_2+
  \frac{v_2}{R_2})^2 \right].    \nonumber
  \end{eqnarray}
Here we have introduced a short notation for components of the
  velocity and of the vector $\dot{\vec{m}}$:
  \begin{equation}
  v_i = \dot{\vec{X}}_{0} \vec{e}_i,  \;\;\;\dot{m}_i = \dot{\vec{m}}\vec{e}_i
  = -\vec{m}\dot{\vec{e}}, \;\;\; i=1,2; \;\;\; v \equiv
   \dot{\vec{X}}_0 \vec{m}.
   \end{equation}
   When the conditions (38) are satisfied, in formula (45) one should
   put $v_i=0$.

   Let us present some implications of the formula (45). First, if each
   piece of the core is at rest at certain instant of time,
   then (at this moment)
$v_i =0 = \dot{m}_i$, and therefore formulae (44),(45) give
\begin{equation}
\frac{ \xi_0}{2 l_0} \approx 1+ 2 l_0^2 \left( \frac{1}{R^2_1}
+\frac{1}{R^2_2} +\frac{1}{R_1 R_2} \right).
\end{equation}
Thus, bending the domain wall is associated with making it thicker
(if we do not  excite the oscillatory component).

If a piece of the core has a nonzero velocity at certain instant of time,
then the terms in the square bracket present on the r.h.s. of formula (45)
may compensate the effect of non-zero curvature because of the minus sign.
To check this possibility, we have considered the following class of exact
solutions of Eq.(24):
\begin{equation}
\vec{X}(x^1,x^2,\tau) = \left(
\begin{array}{c}
x^1 \\
x^2 \\
f(l_1 x^1+l_2 x^2-\tau)
\end{array}
\right),
\end{equation}
where $l_1,l_2$ are constants obeying the condition $ l_1^2+l_2^2=1$, and $f$
can be any smooth function. This plane wave type solution gives
a membrane "levitating" over the $(x^1,x^2)$ plane on the altitude
$x^3=f(l_1 x^1+l_2 x^2-\tau)$, with an infinite plane wave of constant
shape  propagating as a whole along the membrane in the
 direction $(l_1,l_2)$ with the
velocity of light. The fact that such solutions exist can be deduced from
a particular domain wall solution found in \cite{20}. Straightforward
computations give  for the solution (48)
\begin{equation}
\frac{1}{R_2} =0,\;\; \frac{1}{R_1} = \frac{f''}{(1+f'^2)^{3/2}},\;\;
v=-\frac{f'}{\sqrt{1+f'^2}},  \nonumber
\end{equation}
\begin{equation}
\dot{m}_1= \frac{f''}{1+f'^2},\;\;
\dot{m}_2=0,\;\;\;\; v_1=-\frac{f'^2}{\sqrt{1+f'^2}},
\;\; v_2=0.  \nonumber
\end{equation}
It follows from formula (45) that in this case $N^{(1)}=0$. Thus,
in the leading order the
presence of the plane wave travelling along the domain wall does not
influence the width of the domain wall.

\vspace*{0.5cm}
\section{The energy of the domain wall}

The total energy and energy density  are very important
physical characteristics of a solution of field equations, and therefore we
would like to calculate them for our domain wall solutions. The  form
of expression for the total energy depends on whether one integrates over the
hypersurface of constant  $\tau$, or over the hypersurface of constant
laboratory  time $x^0$. The construction of the conserved energy-momentum in
the former
 case has been presented in  Section V of the paper \cite{17}. In the
 present paper we consider the standard energy, i.e. the one obtained
 by integration over the hyperplane of constant $x^0$.

Because the total energy of the field is constant during
 the motion of the domain wall, we
can calculate it at an arbitrarily chosen  laboratory frame
time $x^{0}$. Of course,
in order to really  have  the energy constant in time one should
 use the exact solution
of the field equation (2). It will not come out exactly constant
if we calculate it for an approximate, time dependent
 solution, in particular for the
 one given by formulae  (18).

 The energy-momentum tensor in our model has the following components
 \begin{equation}
 T^{\mu\nu} = \partial^{\mu}\Phi \partial^{\nu}\Phi + \eta^{\mu\nu} L,
 \end{equation}
 with $L$ given by formula (1). The lab-frame energy $E$ is given by the
 integral
  \begin{equation}
  E=\int d^{3}x T^{00}.
  \end{equation}
  In this formula the field $\Phi$, which is a scalar with respect to
  coordinate transformations, can be regarded as a function of the
  $(\tau,\sigma^1,\sigma^2,\xi)$ variables.
  With the help of a formula for differentiation
  of composite functions  $T^{00}$ can be written in the
   following form
\begin{eqnarray}
\lefteqn{T^{00} = [(M^{-1})^{0a}\Phi_{,a} + n^{0} \Phi_{,\xi}]^{2} } \\
& &  +\frac{1}{2} g_{ab}(M^{-1})^{ac}(M^{-1})^{bd}\Phi_{,c}\Phi_{,d}
+\frac{1}{2} (\Phi_{,\xi})^2 + V(\Phi), \nonumber
\end{eqnarray}
where $a=0,1,2$; $n^{0}$ is the $\mu =0$ component of the
four-vector $(n^{\mu})$; the potential
\begin{equation}
V(\Phi) = \frac{\lambda}{2} (\Phi^2 - \Phi_{0}^{2})^2,
\end{equation}
and the matrix $[M^{ab}]$ (the inverse of $[M_{ab}]$) is given by formula
(16).

Also the volume element $d^{3}x$  is expressed by the co-moving coordinates,
\begin{equation}
d^{3}x =  \sqrt{-g} \frac{h(\tau,\sigma^1,\sigma^2,\xi)}{1-\xi K_{0a} g^{a0}}
d\xi d\sigma^1 d\sigma^2 .
\end{equation}
The integration over the $\xi$ coordinate is effectively restricted to the
interval $(\xi_0, - \xi_1)$, because in the vacuum $T^{00}=0$.
Observe that on the hyperplane of constant $x^{0}$ the $\tau$ variable
in general becomes a function of $\xi$; this is because
\begin{equation}
\tau + \xi n^{0}(\tau,\sigma^{i}) = x^{0}=constant.
\end{equation}
Thus, constant $x^{0}$ does not correspond to constant $\tau$,
 except for the core where $\xi=0$ and $x^{0}=\tau$. This complicates
 very much calculation of the integral over $\xi$ because in general we do not
 know the explicit form of the $\tau$ dependence -- for this one would
 have to know explicit solutions of Eq.(24).

There is a particular case in which the calculation of the energy is
relatively simple: when each piece of the core  of the domain wall is
 at rest at the  initial time $x^0$.
Mathematically, this means that  the first derivative of $\vec{X}$
 with respect to $x^0$ vanishes. Then, one can show that
 $n^{0}=0$, hence $x^0 = \tau$ for all $\xi$.

In the following part of this Section we will
use the physical coordinates defined by formula (34) and the gauge
condition (38). In these coordinates $d\sigma^1 d\sigma^2 = |dS|$, the area
of the infinitesimal element $dS$ of the core.
 For the core at rest $v=0$ and $\dot{\vec{e}}_{i} =0$. This implies
that
\begin{equation}
d^{3}\vec{x} = (1-\frac{\xi}{R_1})(1-\frac{\xi}{R_2}) d\xi |dS|,
\end{equation}
and
\begin{eqnarray}
\lefteqn{T^{00} =
  \frac{1}{2[1+\xi(\frac{1}{R_1}+\frac{1}{R_2})]^2} (\Phi_{,\tau})^2 }  \\
& &   + \frac{1}{2 (1-\frac{\xi}{R_1})^2} (\Phi_{,\sigma^1})^2
+ \frac{1}{2 (1-\frac{\xi}{R_2})^2} (\Phi_{,\sigma^2})^2
  + \frac{1}{2} \Phi_{,\xi}^2 +  V(\Phi). \nonumber
\end{eqnarray}
Here we have also used the relation (40). Therefore, formulae (57),
(58) are valid for the Nambu-Goto domain walls only.

 The next step is to integrate $T^{00}$ over $\xi$ in the interval
$(\xi_0, -\xi_1)$. The integrand is a rational function of $\xi$. There is
no danger of a vanishing denominator, because we have assumed that the width
of the domain wall ($2\xi_0$) is smaller than the curvature radia $R_i$,
otherwise the region of validity of the co-moving coordinates would  be
too narrow to cover the whole width of the domain wall.

 For $\Phi$ we take the approximate solution (18) with (22), (23) taken
into account: for $-\xi_{0} \leq \xi \leq \xi_{0}$
\begin{equation}
\Phi(\tau,\sigma^1,\sigma^2,\xi) = \frac{3}{2} \Phi_{0} (\frac{\xi}{\xi_0}
- \frac{1}{3} \frac{\xi^3}{\xi_{0}^{3}} ).
\end{equation}
We may write
\begin{eqnarray}
\Phi_{,\tau} \equiv \frac{\Phi_{0}}{\xi_{0}} \frac{\partial\xi_{0}}
{\partial\tau} f_{0}(z),\;\;
 \Phi_{,\sigma^i} \equiv \frac{\Phi_{0}}{\xi_0} \frac{\partial\xi_0}
{\partial\sigma^i}     f_{0}(z),  \\
\Phi_{,\xi} \equiv \frac{\Phi_{0}}{\xi_0} f_{1}(z), \;\;
V(\Phi) \equiv \frac{\lambda}{2} \Phi^4_0 f^{2}_{2}(z) =
\frac{\Phi_0^2}{8l_0^2}
f^2_2(z),         \nonumber
\end{eqnarray}
where $z \equiv \xi/\xi_0$ has values in the
 interval [1,-1], and the dimensionless functions $f_0(z), f_1(z), f_2(z)$
 are given by the following formulae
 \begin{equation}
f_0= - \frac{3}{2} z (1-z^2), \; f_1 = \frac{3}{2} (1-z^2),
\; f_2 = \frac{9}{4} z^2 (1- \frac{1}{3} z^2)^2 - 1.
 \end{equation}

 The energy $E= \int d^{3}x \; T^{00}$ is given by formula
\begin{eqnarray}
E= \frac{1}{2} \Phi_0^2     \int_{the\; core} |dS| \int^1_{-1} dz
     \left\{ \left[\frac{(1-z \frac{\xi_0}{R_1})(1-z \frac{\xi_0}{R_2})}
     {[1+z(\frac{\xi_0}{R_1}+ \frac{\xi_0}{R_2})]^2} (\frac{\partial \xi_0}
     {\partial \tau})^2  \right. \right.    \nonumber \\
\left.      \frac{1-z \frac{\xi_0}{R_2}}{1-z \frac{\xi_0}{R_1}}
  (\frac{\partial\xi_0}{\partial\sigma^1})^2
  +  \frac{1-z \frac{\xi_0}{R_1}}{1-z \frac{\xi_0}{R_2}}
  (\frac{\partial\xi_0}{\partial\sigma^2})^2 \right]
  \frac{1}{\xi_0}  f_{0}^{2}(z)
        \\
 \left.       + \frac{1}{2 l_0} (1-z \frac{\xi_{0}}{R_1})
  (1-z \frac{\xi_0}{R_2})
 \left[\frac{2 l_0}{\xi_0} f_1^{2}(z) + \frac{\xi_0}{2 l_0} f^2_{2}(z) \right]
 \right\},  \nonumber
\end{eqnarray}
where, by the assumption, $\xi_{0}/R_{i} \ll 1$. Let us recall
 that this formula
is valid for domain walls with the core at the instant rest and
obeying the Nambu-Goto
equation (24). Up to this point we have not used  the approximate solutions
for the half-width $\xi_0$ following from the Ansatz (29).

Let us point out two particular consequences of formula (62).
First, we see that all nonuniformities
of the width of the domain wall increase the energy density
  because nonvanishing derivatives $\partial \xi_0 / \partial \sigma^i$ always
  give positive contribution to the integrand.
Also $\tau$-dependence of $\xi_0$ increases the energy.

Second, there is a curvature-dependent energy associated with the width of the
domain wall which is present even in the case of constant width. It is
given by the second term on the r.h.s. of formula (62) ( the one with
$f_1$, $f_2$).

For the solution (59) the dependence on $z$
is explicit, and it is not difficult to perform the integration over $z$.
Because the resulting formula is lengthy we will not present it here
in the general case.
Let us calculate the energy in the particular case,
 when the oscillatory component is absent, the half-width
  $\xi_0$ is constant, and each piece of the core
   is at instant rest.   Then,
 $\xi_0$ is related to the curvatures by formulae (44),(45), with
 $v=v_i=\dot{m}_i=0$.
Because the core is at the instant rest, the curvature radia have
vanishing derivatives with respect to $\tau$, and therefore $\partial \xi_0/
\partial \tau =0$.
 Formula (44) for $\xi_0 / 2 l_0$ is approximate, so we shall calculate $E$
with the same accuracy, i.e. to the second order in $l_0 /R_i$. We obtain
the following result
\begin{eqnarray}
E = \frac{\Phi_{0}^{2}}{2l_0}  \int_{the \; core} |dS| \left[\frac{1}{2}
(c_1 + c_2)
- l_0^2 (c_1 -c_2) (\frac{1}{R_1^2} + \frac{1}{R_2^2} +\frac{1}{R_1 R_2} )
 \right.  \nonumber \\
 \left.    +2 l_0^2 (d_1 +d_2) \frac{1}{R_1 R_2} \right],
\end{eqnarray}
where the constants $c_i, d_i $ are defined as follows
\begin{eqnarray}
c_1 \equiv \int_{-1}^{1} dz f_{1}^{2} (z) = \frac{12}{5},  \;\;
 c_2 \equiv \int_{-1}^{1} dz f_{2}^{2} (z) \approx 0.777,  \nonumber     \\
 d_1 \equiv \int_{-1}^{1} dz z^2  f_{1}^{2} (z) =\frac{12}{35}, \;\;\;
 d_2 \equiv \int_{-1}^{1} dz z^2  f_{2}^{2} (z) \approx 0.066.
 \end{eqnarray}

Immediate consequence of formulae (63),(64) is that domain walls with nonzero
extrinsic curvature can have smaller energy.
This is because $c_1-c_2$ has come out positive.
 For instance, a straight
infinite cylindrical domain wall with the core radius $R$ (in this case
$1/R_1 =0, 1/R_2 = 1/R$) has smaller energy
 per unit area than the planar domain
wall (for which $1/R_1 =0, 1/R_2 =0$). Thus, the domain wall
prefers to have wrinkles. Let us recall that this  has been found
with the help of the approximate solution. It remains to be checked whether
$c_1-c_2$ is positive for the corresponding exact solution of Eq.(2).
 For this reason we do not claim that this is a proven result -- it is
 just an indication, to be checked in another investigation.

Another interesting problem, namely finding local minima
 of the energy $E$, e.g. for fixed area $|S|$ of  a compact core with a
 given genus, we also  leave for a future investigation.

\section{ The axially  symmetrical toroidal domain wall}

In this Section we shall apply the presented formalism to a toroidal
domain wall. Our  motivation for doing this is that
such domain walls are next to planar, cylindrical or spherical ones
 with respect
to complexity of their geometry, and to our best knowledge they have not been
investigated as yet.   Cylindrical or spherical domain walls
  have been considered in, e.g., \cite{21},\cite{9},\cite{13},\cite{17}.
We shall consider the simplest case of a toroidal domain wall,
 characterised by the axial symmetry with respect to rotations around the
 $x^{3}$-axis.

 We shall parametrise the toroidal core of our domain wall (at each instant
 of time) by two angles, $\phi\in[0,2\pi]$ and $\theta\in[0,2\pi]$:
 \begin{equation}
 \vec{X}(\tau, \theta, \phi) = \left(  \begin{array}{c}
 ( R(\tau) + r(\tau,\theta) cos\theta )\; cos\phi \\
 ( R(\tau) + r(\tau,\theta) cos\theta )\; sin\phi \\
 r(\tau,\theta)\; sin\theta
 \end{array}
 \right),
 \end{equation}
 where  $\tau$ is the laboratory frame time as introduced by formula (4),
 $\phi$ is the azimuthal angle in the $(x^{1},x^{2})$ plane, and $\theta$
 is the angle parametrising cross sections $C$ of the torus with the
half-planes
of constant     $\phi$. $C$ does not have to be a circle.
Because of the axial symmetry, the cross sections C are identical for all
angles $\phi$.
The radius $R(\tau)$ gives the distance from the $x^{3}$-axis
to a "central" circle of the torus. Only if the cross sections $C$
 are circular there is a natural choice for $R(\tau)$:
the distance from the $x^{3}$-axis to the centers of the circles $C$.
In general, there is a freedom in the choice of the "central" circle; this
circle is merely a mathematical construct  -- the physical object is
the domain wall. The choice of $R(\tau)$ has influence on the form of the
$r(\tau,\theta)$ function. In the following we choose
\begin{equation}
R(\tau) = constant \equiv R_{0}.
\end{equation}
Thus, we have to calculate only one function $r(\tau,\theta)$. For
correctness of the parametrisation (65) it is required that
$r(\tau,\theta) > 0$ and $R_{0} > r(\tau,\theta)$. It might happen that
 in order to follow evolution of the torus for
a prolongated interval of time  it is necessary to introduce
several parametrisation patches given by (65) with different $R_{0}$'s.
This is the case when after some time the
 "central" circle with the fixed radius
$R_{0}$ turns out to lie outside of the displaced torus.

Inserting the Ansatz (65) into Eq.(24) we obtain after
 straightforward computations
the following nonlinear equation for $r(\tau,\theta)$:
\begin{eqnarray}
r (r^2 + r'^{2}) \ddot{r} - 2 \dot{r} r' (\dot{r}' r - \dot{r} r')
+ (1-\dot{r}^{2}) (r^2 + 2 r'^{2} - r r'')  \nonumber \\
+ (r^2 + r'^2 - \dot{r}^2 r^2) (1 - \frac{R_{0} -
 r' sin\theta}{R_{0} + r cos\theta}) =0,
\end{eqnarray}
where $\dot{r} \equiv \partial r(\tau,\theta) / \partial \tau$, $r' \equiv
\partial r(\tau,\theta) / \partial \theta$.

In general, equation (67) is not identical with formula (40) -- they
are written in different coordinates.  The point is
that $\vec{X}$ given by formula (65), when expanded in $\theta-\theta_0,
\phi-\phi_0$      in a
 vicinity of a point $(\theta_{0},
\phi_{0})$, does not have the form (34), and
also conditions (38) are not satisfied, in general.
In other words, reparametrisation gauge fixing implied by the formulae
(65), (66) is  different from the gauge fixing implied by
 the formula (34) and condition (38). Equation (67) can of course be
 transformed to the form (40), but this transformation involves nontrivial
 changes of coordinates.

For simplicity, in the following part of this Section we shall consider
only tori which at certain initial instant $\tau_0$ have circular cross
sections. Such tori, in addition to the axial symmetry have also a reflection
symmetry $x^3 \rightarrow -x^3$.   We also assume that for
 $\tau = \tau_0$ each  point of the torus has zero
velocity.  This case is sufficient in order to get
a feeling about evolution of the toroidal domain walls.

 One may ask whether the initially circular cross section will stay
  circular at later times. To investigate this, we  recalculate the
 basic equation (24) for the Ansatz (65) without the assumption (66) that
 $R$ is constant in time. If the cross section $C$ stays circular, there
 should exist such a choice of the function $R(\tau)$ that the solution
  $r$ is constant in $\theta$.
 We have found that such a choice is not possible. Therefore, the
 initially circular cross section will always be deformed during evolution of
the
 torus.

Qualitative picture of the motion governed by Eq.(67) can be
 obtained by considering
motion of the outermost and innermost circles on the core,
 $\theta = 0$ and $\theta = \pi$, respectively. Because of the
symmetries of the torus, $r' = 0$ for $\theta =0,\; \pi$, for all times.
 Therefore, for
$\theta = 0$ Eq.(67) reduces to
\begin{equation}
\frac{\ddot{r}}{1 - \dot{r}^2} = - (\frac{1}{r} - \frac{r''}{r^2}) -
\frac{1}{R_{0} + r}.
\end{equation}
Similarly, for $\theta = \pi$ we have
\begin{equation}
\frac{\ddot{r}}{1 - \dot{r}^2} = - (\frac{1}{r} - \frac{r''}{r^2})
 + \frac{1}{R_{0} - r}.
 \end{equation}

One can check that  equations (68), (69) coincide with formula (40). The main
   radia of curvature of
 the torus at the point $(\theta, \phi)$ are given by the formulae
 \begin{equation}
 \frac{1}{R_{1}} =   \frac{1}{\sqrt{r^2 + r'^2}}
 \frac{r' sin\theta + r cos\theta}
 {R_{0} + r cos\theta},  \nonumber
 \end{equation}
 \begin{equation}
 \frac{1}{R_{2}} = - \frac{r r'' - 2 r'^{2}
  - r^2}{(r^2 + r'^{2})^{3/2}}. \nonumber
 \end{equation}
 Because $r'=0$  for $\theta = 0,\; \pi$, the curvature radia are,
 respectively,  $R_{1}=\pm (R_{0} \pm r)$,
 $\;R_{2}^{-1} =  1/r - r''/r^2\;$
 (we choose the convention that the normal $\vec{m}$ to the torus is
 directed inwards). Also $\dot{\vec{X}}$ does not have the tangent components
 for $\theta = 0,\; \pi$, and $v=-\dot{r}$ in the both cases.
 Now it is easy to check that Eqs.(68), (69)
follow directly from formula (40).

 Equations (68), (69) imply that there are two classes  of  tori,
 differing by their motion.
To see this, let us consider Eqs.(68),(69) at the initial
 time $\tau_0$.
 Then $C$ is a circle. Let us denote its radius by $r_0$. It is clear
 that $r'_0 = r''_0=0$.
Also, by the assumption, $\dot{r}=0$ at the initial time.
 Therefore, on the r.h.s. of Eq.(68) we have the force
 \[-\frac{1}{r_{0}} - \frac{1}{R_{0}+r_{0}}, \]
  which is always negative. It follows that the outermost
 circle "$\theta = 0, \phi - variable$" will always move towards the
 $x^{3}$-axis. As for Eq.(69), the force is equal to
 \[ -\frac{1}{r_0} + \frac{1}{R_{0} - r_{0}}, \]
  and it is negative only for $r_{0} < R_{0}/2$.
 For $r_{0} = R_{0}$ it vanishes, while for $r_{0} > R_{0}/2$ it is positive.
 Therefore, in this last case the circle $\theta = \pi$ will start to
 move away from the $x^3$-axis. To summarize, when the initial
 circular cross section $C$ of the torus is small
  enough, the torus will shrink
 towards a circle in the ($x^1, x^2$)-plane with the center on the $x^3$-axis.
 On the other hand, if the initial circular cross section is sufficiently
 large, we expect that the torus will start to shrink
  towards the $x^3$-axis. As
 for the intermediate case of $r_{0} = R_{0}/2$, initially the acceleration of
 the innermost circle "$\theta = \pi, \phi - variable$" vanishes. At later
times, as
 the cross section $C$ of the torus shrinks $r''$ becomes negative
 (because $r$ has now a local maximum at $\theta = \pi$), and
on the r.h.s. of Eq.(69) the negative term $r''/r^{2}_{0}$ will appear.
 Therefore, we expect that in the case of initial data such that
 $r_{0} = R_{0}/2,\; v_{0}=0$, the circle $\theta = \pi$ will start to move
away
 from the $x^3$ axis.
 Numerical solutions of equation (67) confirm this expectations, see
 Figs. 1a$\div$1c.

 Till now we have discussed evolution of the core of the toroidal domain
 wall. From formulae (44), (45) we obtain approximate value
  of the width of the toroidal domain wall in the absence of the oscillatory
 component.
This is especially simple for $\theta=0, \pi$, because there $\dot{\vec{m}}=0$
and $v_1=v_2=0$.
 We find that
\begin{equation}
\frac{\xi_0}{2 l_0} \approx 1+ \frac{2l_0^2}{1-v^2}\left[\frac{1}{r^2}
(1-\frac{r''}{r})^2
+\frac{1}{(R_0 \pm r)^2}  \pm \frac{1}{r(R_0 \pm r)}\right], \nonumber
\end{equation}
 where the $\pm$ signs are for $\theta=0,\pi$, correspondingly.
{}From these formulae one can see that for $r/R_0$ small enough (a slim ring
far
away from the $x^3$-axis) the outer side ($\theta=0$) of the toroidal
domain wall is thicker than
the inner side ($\theta=\pi$). On the other hand, if
$r/R_0$ becomes closer to 1
 (a fat ring close to the $x^3$-axis) then, vice versa, the inner side is
thicker
 than the outer one (when the oscillatory component $\Omega$ is absent).

As for evolution of the symmetric toroidal domain wall at later times,
 when the rest-frame curvature radia become too small for the Ansatz (18)
to be a reasonable approximation, one can make some guesses based on results
of numerical computations, \cite{13}. We expect that the slim
torus will completely disperse
into radiation after the first collapse to a circle in the $(x^1,x^2)$-plane.
The fat torus will first change its topology from  toroidal to
spherical one --
 this will happen when the innermost circle of the torus is shrunk to the
origin $x^1=x^2=x^3=0$. In the process some energy will be lost into radiation.
Next, the domain wall will evolve as a deformed sphere. Eventually it will
collapse to the origin, and it will disperse into radiation.
This scenario might be different if we pass to another model.
 As pointed out in \cite{13},
sine-Gordon domain walls can reemerge after the collapse. In this case the
domain walls can pass through each other with only a partial
 loss of energy into the radiation, and therefore one can expect
 that the torus  collapses and reemerges several
  times before  it follows the previous  scenario in a final collapse.
 It has also been noticed in the paper \cite{13} that cylindrical sine-Gordon
 domain walls do not bounce in contradistinction to the spherical ones.
This suggests that whether the  sine-Gordon torus bounces or disappears
 after the first collapse
might also depend on the big radius of the torus. At the initial time it is
equal to $R_0$, and the curvature $1/R_1$ depends on it,
see formula (70). For large $R_0$ the torus locally is like a  straight
cylinder, while for small $R_0$ curvature is more pronounced and such
torus might   behave more like a spherical domain wall.
Probably the only way to verify these scenarios is to perform numerical
calculations.

\section{ Remarks}

(a)  Calculations presented in Section 2 of this paper show that the
method proposed in \cite{17} for cylindrical
 and spherical domain walls can be
applied also in the more general case. The resulting equation (24)
for the core of the domain wall is equivalent to the Nambu-Goto equation
(25). Our method seems to be quite universal. Change in the basic equation
(2) would result only in different form of the reccurence relations (19),(20).
For example, if we change the quartic potential $V$ (formula (54)) to the
sine-Gordon type then only numerical coefficients in (19),(20) are different.
The polynomial approximation can also be applied to
 non-relativistic domain walls in condensed matter physics.

We have not obtained any terms with higher derivatives
 in the equation for the
core. Such terms are usually related to so called extrinsic
curvature corrections in the effective action for the core, \cite{9}.
Nevertheless, we would not conclude that the pure
Nambu-Goto equation for the core is the exact, final answer for all
domain walls  with small curvature, for two reasons. First, our solution
is approximate one, and a better approximation could reveal corrections to
the Nambu-Goto equation. Second,
 the basic field equation (2) possesses infinitely many
solutions  in the topological class of the single domain wall. Our cubic
polynomial Ansatz probably picks  only a subset of
 these solutions. There might be other
solutions for which the core would not obey the pure Nambu-Goto equation (25).

Actually, it is easy to point out
  possible ways to improve our approximation and  to find  more general
domain wall  solutions.
   Natural step to find a wider class of solutions within the cubic
polynomial approximation
  is to  divide the interval $(\xi_{0}, - \xi_{1})$ (see
   formula (18)) into two
 subintervals, and to use  two independent cubic polynomials in each of
 them. Next, one should match smoothly the two polynomials with each other
  and with the vacuum  fields to obtain continuous $\Phi$ and $\partial_{\xi}
\Phi$.
 In order to improve our approximation
 one could use polynomials of  higher order, as suggested by
 considerations presented in the Appendix.   Work along these lines
 is in progress.

(b)  Formula (40) from Section 3, giving the transverse acceleration
in terms  of the local curvature radia, is very useful in
 providing qualitative
  understanding of time evolution of the core. For example, one could
  foresee from it that a bump on otherwise almost flat core will
  first split into two bumps travelling in the opposite directions,
  next more smaller bumps will appear and finally the bump will
  disperse into small riples. Numerical computations we have carried out
  in the case of an axially symmetric bump on the toroidal core
  described by equation  (67) confirm this expectation.

(c)  Let us end this Section by mentioning several possibilities
  to extend our work. The first one is  to improve our description of
  dynamics of the domain walls by including possible radiation,  creation
  of pairs "domain wall + anti-domain wall", and interactions between
  close parts of the domain wall.

Second, one would like to have an analytical (approximate) description
of domain walls also in the case when the curvature radia of the core are
comparable with the width of the domain wall. Our results are for the case
when the radia are much larger than the width.

   Third, one could try to apply
    the polynomial approximation to curved, non-static
  vortices in relativistic field theories as well as in condensed matter
  physics. We have seen that this approximation is capable of yielding
  rather detailed information about dynamics of the width of the domain wall.
It would be very interesting to have such an information also for vortices.

\vspace*{1cm}

{\bf THE APPENDIX. The accuracy of the polynomial approximation}

Here we would like to comment on accuracy of the approximation used in
this paper. It is  instructive  to apply the above presented scheme
to the planar domain wall given by the exact solution (3).
Then,
\[ K_{ab} =0,   \;\; (g_{ab}) = diag(-1,1,1), \;\; \xi = x^3, \]
and in the absence of the oscillating component $\tilde{A}=1$, i.e. $\xi_0
 = 2 l_0$.  Then
 \begin{equation}
 \Phi (x^3) = \Phi^{(1)}(x^3) \equiv \frac{3}{4} \Phi_0\; \frac{x^3}{l_0}
 \; (1 - \frac{1}{12} (\frac{x^3}{l_0})^2).  \nonumber
 \end{equation}
 The boundaries of this domain wall are at $x^3 = \pm 2 l_{0}$. Comparing
 $\Phi^{(1)}$ with the solution (3), we see that it is not equal to
  the first two terms
 of Taylor expansion of the solution (3). Thus, we do not recover the exact
solution
 (3) term by term in the Taylor expansion. On the other hand, the approximate
 solution $\Phi^{(1)}$ has the right global characteristics of the domain
 wall like energy per unit area or the boundary conditions.
 The energy per unit  area in the case of solution (3) is
 \[ E_0= \frac{2}{3} \frac{1}{l_0} \Phi_{0}^2.   \]
 It is minimal in the topological class of  single domain
 wall, \cite{22}.
 The energy of the solution $\Phi^{(1)}$ is a little bit higher
 \[ E^{(1)} \approx 0.79 \frac{1}{l_0} \Phi_{0}^2.  \]
For a comparison of  energy densities,  see Figs.2a,b.

Improving the approximation (73)  consists of
 including terms of higher order in $x^3$.
We know that the exact solution
is an odd function of $x^3$, so  the first term to be included
is proportional to the fifth power of $x^3$. However,
it turns out that a fifth order polynomial in $x^3$ can not
simultaneously obey recurrence relations obtained from Eq.(2) and
the continuity conditions
for $\Phi$ and $\partial_{x^3}\Phi$. The reason is that the
recurrence relations imply
that the  fifth power of $x^3$ comes with a positive coefficient;
then it is easy to see that there can be a problem with smooth matching
of the polynomial with the constant vacuum solutions.

Going to the seventh order polynomial does
   yield a solution. Denoting $z \equiv x^3/l_0$,
the result we have obtained can be written in the form
\[ \Phi^{(2)} \approx \Phi_0 \;(0.5750 z - 0.0479 z^3 + 0.0059 z^5
- 0.0004 z^7).   \]
$\Phi^{(2)}$ smoothly  matches the vacuum solutions for $x^3 \approx
 \pm 2.593 l_0$,
while for the solution $\Phi^{(1)}$ the matching takes place at $x^3 = \pm 2
l_0$.
Energy per unit area for the solution $\Phi^{(2)}$ is equal to
\[ E^{(2)} \approx 0.706 \frac{1}{l_0} \Phi^2_0, \]
which higher than $E_0$ by only 6\%.
 Also the energy densities do not differ
much from their exact values, see Figs.3a,b.

To summarize,  the direct comparison
presented above in the case of the planar domain wall
 shows that the polynomial approximation
 works rather well. Therefore, we expect that also for slightly curved
 domain walls,  i.e. such that their rest-frame curvature radia are
large in comparison with their width,
the cubic polynomial approximation is a reasonable one, and that  in order
to improve it one could  just use a higher order
 polynomial. Calculations with seventh order polynomials do
  not have to be cumbersome,
 because Taylor expansions and other necessary operations
 can be carried out by a computer.

\newpage
\begin{centerline}
{\bf Figure captions}
\end{centerline}
\vskip 12pt
\hspace*{-6mm}Fig.1a. Numerical solution of equation (67).
The initial data are $r(0,\theta)=0.4,\;\; \dot{r}(0,\theta)=0$.
The contours are the cross sections C of the torus with,
 e.g., $(x^1,x^3)$ plane. They are determined  by the function
$r(t,\theta)$ (with $\theta$ changing from 0 to $2\pi$). The cross sections
are shown for the following values of
 $t \equiv \tau/R_{0}$: $0.00,\;0.15,\;0.25,\;0.35,\;0.45,\;0.55$.
  The horizontal axis shows the distance from the $x^3$ axis, measured in
  the unit $R_{0}$.
\vskip 12pt
\hspace*{-6mm}Fig.1b. The same as in Fig.1a, but with the initial data
$r(0,\theta)=0.5$, $\dot{r}(0,\theta)=0$. The values of $t$ are: $0.00,\;
0.15,\;0.25,\;0.35,\;0.45,\;0.55,\;0.65.$
\vskip 12pt
\hspace*{-6mm}Fig.1c. The same as in Fig.1a, but with the initial data
$r(0,\theta)=0.65$, $\dot{r}(0,\theta)=0$. The values of $t$ are: $0.00,\;
0.15,\;0.25,\;0.35,\;0.45,\;0.55,\;0.65.$
\vskip 12pt
\hspace*{-6mm}Fig.2a. The comparison of the gradient contributions
$1/2 (\partial _{x^3}\Phi)^2$ to the energy density of the planar domain wall.
The solid line corresponds to the approximate solution $\Phi^{(1)}$,
 the dashed line corresponds to the exact solution (3).
\vskip 12pt
\hspace*{-6mm}Fig.2b. The comparison of the potential energies $V(\Phi)$
(formula (54)) for the planar domain wall. The solid line corresponds
to the approximate solution $\Phi^{(1)}$, the dashed line corresponds to
the exact solution (3).
\vskip 12pt
\hspace*{-6mm}Fig.3a.  The same as in Fig.2a, but for the approximate
solution  $\Phi^{(2)}$.
\vskip 12pt
\hspace*{-6mm}Fig.3b. The same as in Fig.2b, but for the approximate
solution $\Phi^{(2)}$.

\end{document}